\documentclass[a4 paper,12 pt]{article}

\begin{document}
\title{\bf{Color Dielectric Model with two Scalar Fields}}
\author{A. Wereszczy\'{n}ski $^{a)}$ \thanks{wereszcz@alphas.if.uj.edu.pl} \, and
{M. \'{S}lusarczyk $^{a,b)}$ \thanks{mslus@phys.ualberta.ca}}
       \\
       \\ $^{a)}$ Institute of Physics,  Jagiellonian University,
       \\ Reymonta 4, Krakow, Poland
       \\
       \\  $^{b)}$ Department of Physics, University of Alberta,
       \\ Edmonton, Alberta T6G 2J1, Canada}
\maketitle
\begin{abstract}
$SU(2)$ Yang-Mills theory coupled in a non-minimal way to two
scalar fields is discussed. For the massless scalar fields a
family of finite energy solutions generated by an external,
static electric charge is found. Additionally, there is a single
solution which can be interpreted as confining one. Similar
solutions have been obtained in the magnetic sector. In case of
massive scalar fields the Coulomb problem is investigated. We find
that asymptotic behavior of the fields can also, for some values
of the parameter of the model, give confinement of the electric
charge. Quite interesting one glueball--meson coupling gives the
linear confining potential. Finally, it is shown that for one
non-dynamical scalar field we derive the color dielectric
generalization of the Pagels--Tomboulis model.
\end{abstract}
\newpage
\section{\bf{Introduction}}
There exists several attempts to understand dynamics of gluonic fields in
non-perturbative region. Among them color dielectric models seem to be
especially promising \cite{cdiel1}. Although none of them describes
the full range of observable phenomena in a satisfactory way, they are simple
enough to carry out an extensive study of classical solutions and provide
guidelines for constructing correct effective theory for non-perturbative
QCD.

A class of color dielectric-like models was proposed by Dick in the late 90's.
It was inspired by low--energy limit of certain string theories. In the simplest case,
with one scalar field coupled non-minimally to the $SU(2)$ gauge field, Dick models
reproduce a broad family of quark--anti-quark confining potentials. As it has been
pointed out  in our previous papers \cite{my1} one can easily fit potentials
originating from these models to phenomenological ones, obtained from heavy quarkonia
spectroscopy.

Correct potential between quarks is of course only one in the whole family of
factors to be considered and reproduced by an effective gluodynamics theory.
Acceptable model should also exclude colored states from physical spectrum, reproduce
energy--momentum tensor trace anomaly as well as predict existence
of massive particles build entirely of gluonic fields -- glueballs.
The last issue is particularly challenging and was addressed by several authors in
the last few months (cf. eg. \cite{glueballs}, \cite{burakovsky}). Glueball-like interpretation has been
also proposed for dilaton field in the Dick model \cite{dick1, dick2}. However, so far
status of the glueball state in the framework of color dielectric models is not clear.

The most natural generalization of the Dick model emerges when dilaton field is
replaced by a pair of two scalars. Theory of this kind was recently studied by
Bazeia {\it at al.} \cite{bazeia} in the context of confining potential in a
system of two planar domain walls. They concluded that a theory with two scalars
is generally better suited to description of anti--screening effects. As we will show
below, color dielectric model with two scalars possesses also other striking features.
Surprisingly observed effects are not a simple superposition of
well--known results for one scalar.  In particular, non--minimal two scalar--gauge
coupling has glueball as well as meson interpretation and simultaneously
gives linear confining potential for external sources. For non-dynamical scalar
fields it is also equivalent to the well--known Pagels--Tomboulis model and its
generalized version, which correctly reproduces the trace anomaly in the language of
effective models.

The plan of this paper is the following. In the next section we
introduce and shortly justify the most general form of the model
discussed. In section \ref{cou1} standard analysis of Coulomb
problem for external source in the case of massless scalar field
is carried out which eventually leads to confining behavior. It also
allows us to obtain bounds on the parameters and rule out the whole
family of models which do not lead to correct confining potential.
Afterwords we deal with magnetic monopoles and discuss their possible
application to glueball physics. In section \ref{massec} we present
similar discussion for massive scalars.
Finally, non--dynamical scalars are studied in section \ref{nondynamical}.
%
\section{The model}
%
In the present paper we focus on the following density of the
Lagrange function
\begin{equation}
L=-\frac{1}{4}\sigma( \phi, \psi ) F^{c \mu \nu }F_{\mu \nu }^c
+\frac{1}{2} \partial_{\mu } \phi \partial^{\mu } \phi
+\frac{1}{2}
\partial_{\mu } \psi \partial^{\mu } \psi -\frac{1}{2} m_{\phi}^2 \phi^2
- \frac{1}{2} m^2_{\psi} \psi^2 -V(\phi, \psi ), \label{model}
\end{equation}
where the coupling between scalar and gauge fields is chosen to be
\begin{equation}
\sigma( \phi, \psi)= \left( \frac{\phi}{\Lambda } \right)^a \left(
\frac{\psi}{\Lambda } \right)^b. \label{diel}
\end{equation}
Here $\Lambda $ is a dimensional constant: $[\Lambda] = cm^{-1}$ and $a, b$ are positive
parameters. A particular form of the potential $V(\phi ,\psi )$,
fixing asymptotic value of the scalar fields will be specified
later.

We would like to notice that all analytical solutions obtained
below can be also found for the more general coupling function
\begin{equation}
\sigma (\phi, \psi )= \frac{ \left( \frac{\phi}{\Lambda }
\right)^a \left( \frac{\psi}{\Lambda } \right)^b }{1+\left(
\frac{\phi}{\Lambda } \right)^a \left( \frac{\psi}{\Lambda }
\right)^b}. \label{dielgeneral}
\end{equation}
Due to the fact that for $\phi \rightarrow \infty, \,  \psi
\rightarrow \infty $ the gauge part of the model takes the
standard Yang--Mills form i.e. $\sigma \rightarrow 1$, the short
range behavior of the fields changes. For instance, the electric
field falls as $r^{-2}$ for $r \rightarrow 0$. One can check that
in the presented solutions the additional, well-known
Coulomb term will appear. However, such form of the coupling does not
modify the most interesting long range behavior of fields.
Thus we will neglect the denominator in the coupling function that
is restrict discussion to the dielectric function given by
(\ref{diel}).
\section{\bf{The massless scalar fields}}
\label{cou1} Firstly we analyze the simplest case where the scalar
fields are massless. Additionally, the potential term is
neglected. Due to that, as we will show it in the next subsection,
the asymptotic values of the scalar fields can be arbitrarily
large. These values correspond to the so-called dilaton and
modulus charge \cite{cvetic}. The Lagrange density has the
following form
\begin{equation}
L=-\frac{1}{4} \left( \frac{\phi}{\Lambda } \right)^a \left(
\frac{\psi}{\Lambda } \right)^b F^{c \mu \nu }F_{\mu \nu }^c
+\frac{1}{2}
\partial_{\mu } \phi \partial^{\mu } \phi +\frac{1}{2}
\partial_{\mu } \psi \partial^{\mu } \psi. \label{model1}
\end{equation}
The pertinent equations of motion read
\begin{equation}
D_{\mu } \left( \left( \frac{\phi}{\Lambda } \right)^a \left(
\frac{\psi}{\Lambda } \right)^b F^{c \mu \nu } \right)=j^{c \nu }
\label{eqmot1}
\end{equation}
and
\begin{equation}
\partial_{\mu } \partial^{\mu } \phi =-\frac{a}{4\Lambda} \left( \frac{\phi}{\Lambda } \right)^{a-1}
 \left(\frac{\psi}{\Lambda } \right)^b F^{c \mu \nu }F_{\mu \nu }^c
\label{eqmot2}
\end{equation}
\begin{equation}
\partial_{\mu } \partial^{\mu } \psi =-\frac{b}{4\Lambda} \left( \frac{\phi}{\Lambda } \right)^{a}
 \left(\frac{\psi}{\Lambda } \right)^{b-1} F^{c \mu \nu }F_{\mu \nu
 }^c,
\label{eqmot3}
\end{equation}
where $j^{c \nu }$ is an external current.
\subsection{\bf{Electric solutions}}
\label{cou2}
Let us now investigate the Coulomb problem i.e. we will find field
configuration generated by an external, static, point-like
electric source:
\begin{equation}
j^{c \mu }=4\pi q \delta (r) \delta^{c3} \delta^{\mu 0}.
\label{current}
\end{equation}
Moreover, for simplicity, the source is set to be Abelian. We
would like to stress that restriction to the Abelian case is not
essential. One can easily analyze the more general source with
three non-zero color components. However, the results will be
modified only by a multiplicative color-dependent constant (cf.
\cite{dick1}). The dependence on spatial coordinates is identical.
Because of that we remain in the Abelian sector. In order to work
only with Abelian degrees of freedom one can put
$$A_{\mu }^c=A_{\mu } \delta^{c3},$$ and field equations can be
rewritten as follows:
\begin{equation}
\left[ r^2 \left( \frac{\phi}{\Lambda } \right)^a \left(
\frac{\psi}{\Lambda } \right)^b E \right]'=4\pi q \delta(r),
\label{eqmotless1}
\end{equation}
and
\begin{equation}
\nabla^2_r \phi = -\frac{aE^2}{2\Lambda} \left(
\frac{\phi}{\Lambda } \right)^{a-1}
 \left(\frac{\psi}{\Lambda } \right)^b,
\label{eqmotless2}
\end{equation}
\begin{equation}
\nabla^2_r \psi = -\frac{bE^2}{2\Lambda} \left(
\frac{\phi}{\Lambda } \right)^{a}
 \left(\frac{\psi}{\Lambda } \right)^{b-1}.
\label{eqmotless3}
\end{equation}
Here $E^{ci}=-F^{c0i}$ and $\vec{E}^c=\vec{E} \delta^{c3}$ i.e.
the electric field points in the same color direction as the
source. We have also assumed spherical symmetry of the problem
$\vec{E}=E(r)\hat{r}$. The prime stands for differentiation
with respect to $r$. Using the Gauss law (\ref{eqmotless1}) one
can express the electric field in terms of the scalar fields
\begin{equation}
E = \frac{q}{r^2} \left( \frac{\phi}{\Lambda } \right)^{-a}
 \left(\frac{\psi}{\Lambda } \right)^{-b}.
\label{efield}
\end{equation}
One can treat this field as the standard Coulomb field in a very
non-standard medium. Indeed, it can be written as
$E=\frac{q_{eff}}{r^2}$ where $q_{eff}$ is an effective charge
strongly dependent on the scalar fields. Now, we insert it into
the equations for the scalars and get
\begin{equation}
\ddot \phi=-\frac{aq^2}{2\Lambda} \left( \frac{\phi}{\Lambda }
\right)^{-a-1}
 \left(\frac{\psi}{\Lambda } \right)^{-b}
\label{eqmotless4}
\end{equation}
and
\begin{equation}
\ddot \psi=-\frac{bq^2}{2\Lambda} \left( \frac{\phi}{\Lambda }
\right)^{-a}
 \left(\frac{\psi}{\Lambda } \right)^{-b-1},
\label{eqmotless5}
\end{equation}
where a new variable $x=1/r$ has been introduced. Here dot denotes
differentiation with respect to $x$. We look for solutions of the
equations (\ref{eqmotless4}) and (\ref{eqmotless5}) in the
power-like form. Namely:
\begin{equation}
\phi (x) =A\Lambda \left(\frac{x}{\Lambda}+\frac{x_0}{\Lambda}
\right)^n \label{anzatz1}
\end{equation}
and
\begin{equation}
\psi (x) =B\Lambda \left(\frac{x}{\Lambda}+\frac{x_0}{\Lambda}
\right)^m, \label{anzatz2}
\end{equation}
where constants $m$, $n$, $A$, $B$ and $x_0$ are yet to be determined. After
some easy algebra one can find that
\begin{equation}
n=m=\frac{2}{2+a+b} \label{condless1}
\end{equation}
and
\begin{equation}
A=a^{\frac{1}{2+a+b}} \left( \frac{b}{a}
\right)^{-\frac{b}{2(2+a+b)}} \left( \frac{q^2 (2+a+b)^2}{4(a+b)}
\right)^{\frac{1}{2+a+b}}  , \label{condless2}
\end{equation}
\begin{equation}
B=b^{\frac{1}{2+a+b}} \left( \frac{a}{b}
\right)^{-\frac{a}{2(2+a+b)}} \left( \frac{q^2 (2+a+b)^2}{4(a+b)}
\right)^{\frac{1}{2+a+b}}  . \label{condless3}
\end{equation}
Finally, we have found the following family of solutions of the
Coulomb problem labeled by the positive parameter $\beta_0$
\begin{equation}
\phi (r)=A \Lambda \left( \frac{1}{r\Lambda} +\frac{1}{\beta_0 }
\right)^{\frac{2}{2+a+b}} \label{solless1}
\end{equation}
\begin{equation}
\psi (r)=B \Lambda \left( \frac{1}{r\Lambda} +\frac{1}{\beta_0 }
\right)^{\frac{2}{2+a+b}}. \label{solless2}
\end{equation}
The electric field takes the form:
\begin{equation}
E=\frac{q}{r^2} A^{-a}B^{-b} \left( \frac{1}{r\Lambda}
+\frac{1}{\beta_0 } \right)^{\frac{-2a-2b}{2+a+b}}.
\label{solless3}
\end{equation}
Here we have put $\beta_0=\frac{\Lambda}{x_0}$. Then corresponding
energy density reads
\begin{equation}
\epsilon = \frac{1}{2} \left( q^2A^{-a} B^{-b}
+\frac{4(A^2+B^2)}{(2+a+b)^2} \right) \frac{1}{r^4} \left(
\frac{1}{r\Lambda } +\frac{1}{\beta_0 }
\right)^{-\frac{2a+2b}{2+a+b}}. \label{sollessenden}
\end{equation}
Integrating (\ref{sollessenden}) one finds that the total energy of
the fields generated by the external static electric charge is finite
if $a+b >2 $ and it is given by the formula:
\begin{equation}
\mathcal{E}= \Lambda  \frac{a+b+2}{a+b-2} \frac{1}{2} \left(
q^2A^{-a} B^{-b} +\frac{4(A^2+B^2)}{(2+a+b)^2} \right)
\beta_0^{\frac{a+b-2}{a+b+2}}. \label{sollessenergy}
\end{equation}
Of course, there are also solutions with negative parameter
$\beta_0$. However, such field configurations are not regular for
all $r$. There is a singularity at $r \rightarrow \beta $. Due to
that such solutions are usually interpreted as electric black hole
solutions with $r>\beta_0$ \cite{cvetic}.

In case of the family of solutions with finite energy we can define
some additional numbers, so-called dilaton (scalar) and gauge ('electric') charge,
which characterize asymptotic behavior of the solutions.
Using the standard definitions we obtain:
\begin{equation}
Q=r^2 E(r) |_{r \rightarrow \infty}=qA^{-a} B^{-b}
\beta_0^{\frac{2a+2b}{2+a+b}}, \label{adcharge1}
\end{equation}
for the electric charge and:
\begin{equation}
D_{\phi}=-r^2 \frac{d\phi }{dr} \left|_{r \rightarrow \infty}
=\frac{2}{2+a+b} A \beta_0^{\frac{a+b}{2+a+b}} \right. ,
\label{adcharge2}
\end{equation}
\begin{equation}
D_{\psi}=-r^2 \frac{d\psi }{dr} \left|_{r \rightarrow \infty} =
\frac{2}{2+a+b} B \beta_0^{\frac{a+b}{2+a+b}} \right.
\label{adcharge3}
\end{equation}
for the scalar charges.

One can notice that scalar charges and gauge charge are not
independent. They are connected by the simple relation
\begin{equation}
\frac{D_{\phi }D_{\psi }}{Q} = q \frac{a+b}{\sqrt{ab}}.
\label{relcharge}
\end{equation}
The finite energy solutions of the Coulomb problem can be
interpreted as screening configurations. The fields generated by
fixed electric charge can have arbitrarily small energy. This
phenomenon is known from the standard non-Abelian Yang--Mills
theory where non-Abelian contents of the gauge field lowers the
total energy \cite{kiskis}. Here it occurs even in the Abelian
part of the model. This suggests that scalar fields can probably
represent the non-Abelian part of the gluonic sector of QCD
\cite{leupold}.

In addition to the family of the finite energy field
configurations, there is a unique infinite energy solution
\begin{equation}
\phi (r)=A \Lambda \left( \frac{1}{r \Lambda }
\right)^{\frac{2}{2+a+b}}, \label{sollessconf1}
\end{equation}
\begin{equation}
\psi (r)=B \Lambda \left( \frac{1}{r \Lambda }
\right)^{\frac{2}{2+a+b}} \label{sollessconf2}
\end{equation}
and
\begin{equation}
E=q A^{-a} B^{-b} \Lambda^2 \left( \frac{1}{r \Lambda }
\right)^{\frac{4}{2+a+b}}. \label{sollessconf3}
\end{equation}
Then the electric potential reads
\begin{equation}
U=q \frac{a+b+2}{a+b-2} A^{-a} B^{-b} \Lambda \left( \frac{1}{r
\Lambda } \right)^{\frac{2-a-b}{2+a+b}} \label{confpotential1}
\end{equation}
for $a+b \neq 2$ and
\begin{equation}
U= q\Lambda A^{-a} B^{-b} \ln r \Lambda \label{confpotential1a}
\end{equation}
for $a+b=2$. The pertinent energy density is given as
\begin{equation}
\epsilon = \frac{1}{2} \left( q^2A^{-a} B^{-b}
+\frac{4(A^2+B^2)}{(2+a+b)^2} \right) \frac{1}{r^4} \left(
\frac{1}{r\Lambda } \right)^{\frac{-2a-2b}{2+a+b}}.
\label{confdensity1a}
\end{equation}
Obviously, the corresponding total energy is infinite. However,
for $a+b \geq 2$ the infiniteness of the total energy has its
origin in the long range behavior of the fields. It is unlikely
the standard Coulomb potential in the classical electrodynamics
where the energy is infinite because of the singularity of the
potential at $r=0$. This effect has been used in many QCD
motivated theories to model confinement of quarks at the classical
level. The confining electric potential obtained here does not
diverge stronger than linearly with distance $r$. That is in
agreement with the famous Seiler constrains for quark--anti-quark
potential \cite{seiler}. One can observe that the standard linear
potential emerges only in the limit $a+b \rightarrow \infty $
which cannot be implemented at the Lagrangian level. This means
that the model with two scalars does not describe the linear
potential. However, it can be approximated with arbitrary accuracy
by taking sufficiently large value of the parameter $a$ and $b$.
In spite of the fact that our model does not possess the linear
potential, it can still be physically interesting. In fact, it
gives confining potentials which can be compared with many
phenomenological potentials obtained from fits to charmonium and
bottomium states. For example, for $a+b=6$ one obtains confining
part of the Zalewski--Motyka potential \cite{kacper}
\begin{equation}
U_{ZM}= C_1 \left( \sqrt{r} -\frac{C_2}{r} \right), \label{kacper}
\end{equation}
where $C_1 \simeq 0.71 \mbox{Gev}^{\frac{1}{2}}$ and $C_2 \simeq
0.46 \mbox{Gev}^{\frac{3}{2}}$. Analogously, the Martin potential
\cite{martin} is reproduced for $a+b=\frac{22}{9}$.

As long as the asymptotic values of the scalar fields are not
fixed i.e. the model does not contain a potential term for
scalars, the confining and finite energy solutions appear
simultaneously. In order to get rid of the screening field
configurations and to preserve the confining solution one has to
add a particular potential therm. Let us, for instance, choose it
in the following  form, which enables us to obtain analytical
solutions
\begin{equation}
V(\phi, \psi)=\Lambda^4 \sum_{i=1}^N \alpha_i \left(
\frac{\phi}{\Lambda} \right) ^{4+a+b-\beta_i} \left(
\frac{\psi}{\Lambda} \right)^{\beta_i}, \label{potential}
\end{equation}
where every parameter $\beta_i$ fulfills
$0<\beta_i<4+a+b$. Constants $\alpha_i $ are restricted by the
condition that the potential must be positively valued and has
global minimum at $\phi =\psi =0$. Then the equations for the
scalar fields (\ref{eqmotless4}), (\ref{eqmotless5}) can be
rewritten in the form
$$
\phi''=-\frac{aq^2}{2\Lambda} \left( \frac{\phi}{\Lambda }
\right)^{-a-1}
 \left(\frac{\psi}{\Lambda } \right)^{-b} + $$
\begin{equation}
  + \frac{\Lambda^3}{x^4}
 \sum_i  \alpha_i (4+a+b-\beta_i) \left( \frac{\phi}{\Lambda} \right)
^{3+a+b-\beta_i} \left( \frac{\psi}{\Lambda} \right)^{\beta_i}
\label{eqmotless6}
\end{equation}
and
\begin{equation}
\psi''=-\frac{bq^2}{2\Lambda} \left( \frac{\phi}{\Lambda }
\right)^{-a}
 \left(\frac{\psi}{\Lambda } \right)^{-b-1}
 + \frac{\Lambda^3}{x^4} \sum_i  \alpha_i \beta_i \left( \frac{\phi}{\Lambda} \right)
^{4+a+b-\beta_i} \left( \frac{\psi}{\Lambda} \right)^{\beta_i-1}.
\label{eqmotless7}
\end{equation}
The Gauss law remains unchanged and electric field can be expressed by
the scalars via the formula (\ref{efield}). In the simplest case,
when $N=1$ the solutions read
\begin{equation}
\phi (r)=\mathcal{A} \Lambda \left( \frac{1}{r\Lambda}
\right)^{\frac{2}{2+a+b}}, \label{sollessconf4}
\end{equation}
\begin{equation}
\psi (r)=\mathcal{B} \Lambda \left( \frac{1}{r\Lambda}
\right)^{\frac{2}{2+a+b}} \label{sollessconf5}
\end{equation}
and
\begin{equation}
E=q\Lambda^2 \mathcal{A}^{-a} \mathcal{B}^{-b} \left(
\frac{1}{r\Lambda} \right)^{\frac{4}{2+a+b}}, \label{sollessconf6}
\end{equation}
where constants $\mathcal{A}, \mathcal{B}$ are given by the
following set of algebraic equations
\begin{equation}
\frac{2(a+b)}{(2+a+b)^2} \mathcal{A} = \frac{aq^2}{2}
\mathcal{A}^{-a-1} \mathcal{B}^{-b} +\alpha (4+a+b-\beta
)\mathcal{A}^{3+a+b-\beta } \mathcal{B}^{\beta}, \label{constset1}
\end{equation}
and
\begin{equation}
\frac{2(a+b)}{(2+a+b)^2} \mathcal{B} = \frac{bq^2}{2}
\mathcal{A}^{-a} \mathcal{B}^{-b-1} +\alpha \beta
\mathcal{A}^{4+a+b-\beta } \mathcal{B}^{\beta -1}.
\label{constset2}
\end{equation}
One can notice that taking into account the potential
(\ref{potential}) does not influence the functional dependence of
the confining solutions. Modification of the Lagrangian is visible
only in the constants $\mathcal{A}$ and $\mathcal{B}$. Because of
the fact that the potential (\ref{potential}) has unique minimum
for the vanishing scalar fields the finite energy solutions
(\ref{solless1})--(\ref{solless3}) (asymptotically non-zero) can no
longer have finite energy. Moreover, one can check that
such solutions are in contradiction with the equations of motion.
In other words, the finite energy configurations disappear from
our model.
\subsection{\bf{Magnetic monopoles}}
In this subsection we find field solutions generated by the static
magnetic monopole. In order to do it we adopt the well-known $SU(2)$
monopole ansatz
\begin{equation}
 A^{c}_{i} = \epsilon_{cik} \frac{x^{k}}{r^{2}}(g - 1), \; A^{c}_{0} =
0, \label{ansatzmag}
\end{equation}
where $g=g(r)$ is an unknown function. Then the non-Abelian field
equations read
\begin{equation}
\left[ \left( \frac{\phi}{\Lambda} \right)^a  \left(
\frac{\psi}{\Lambda} \right)^b g' \right]' + \frac{1}{r^{2}}
\left( \frac{\phi}{\Lambda} \right)^a  \left( \frac{\psi}{\Lambda}
\right)^b g \left( 1 - g^{2} \right) = 0, \label{eqmonopol1}
\end{equation}
and
\begin{equation}
\nabla^2_r \phi =\frac{a}{2\Lambda } \left[ \frac{2g'^{2}}{r^{2}}
+ \frac{(g^{2} - 1)^{2}}{r^{4}} \right]  \left( \frac{ \phi
}{\Lambda } \right)^{a - 1} \left( \frac{\psi}{\Lambda} \right)^b,
\label{eqmonopol2}
\end{equation}
\begin{equation}
\nabla^2_r \phi =\frac{b}{2\Lambda} \left[ \frac{2g'^{2}}{r^{2}} +
\frac{(g^{2} - 1)^{2}}{r^{4}} \right]  \left( \frac{ \phi
}{\Lambda } \right)^{a} \left( \frac{\psi}{\Lambda} \right)^{b-1}.
\label{eqmonopol3}
\end{equation}
Equation (\ref{eqmonopol1}) possesses the trivial solution
\begin{equation}
g=0. \label{solmon1}
\end{equation}
In other words, we have found the Wu--Yang $SU(2)$ monopole. In the
classical Yang--Mills theory this monopole has infinite energy due
to the singular behavior of the gauge field in the vicinity of the monopole.
As we will see the scalar fields are
able to 'regularize' monopole solution.

After substituting (\ref{solmon1}) to the remaining field
equations we can obtain the following family of solutions
\begin{equation}
\phi = C \Lambda \left( \frac{1}{r\Lambda } +\frac{1}{\beta_0}
\right)^{\frac{2}{2-a-b}} \label{solmon2}
\end{equation}
and
\begin{equation}
\psi = D \Lambda \left( \frac{1}{r\Lambda } +\frac{1}{\beta_0}
\right)^{\frac{2}{2-a-b}} \label{solmon3}
\end{equation}
where the constants read
\begin{equation}
C=a^{\frac{1}{2-a-b}} \left( \frac{b}{a}
\right)^{\frac{b}{2(2-a-b)}} \left( \frac{q^2 (2-a-b)^2}{4(a+b)}
\right)^{\frac{1}{2-a-b}}  \label{constmag1}
\end{equation}
and
\begin{equation}
D=b^{\frac{1}{2-a-b}} \left( \frac{a}{b}
\right)^{\frac{a}{2(2-a-b)}} \left( \frac{q^2 (2-a-b)^2}{4(a+b)}
\right)^{\frac{1}{2-a-b}}  . \label{constmag2}
\end{equation}
The energy density originating from (\ref{solmon1})--(\ref{solmon3}):
\begin{equation}
\epsilon = \frac{1}{2} \left( C^{a} D^{b} +
\frac{4(C^2+D^2)}{(2-a-b)^2} \right) \frac{1}{r^4} \left(
\frac{1}{r\Lambda } +\frac{1}{\beta_0}
\right)^{\frac{2a+2b}{2-a-b}}. \label{monenden}
\end{equation}
As one could expect, for $a+b>2$  the total energy for the field
generated by the Wu--Yang monopole is finite and reads
\begin{equation}
\mathcal{E}= \frac{a+b-2}{a+b+2} \frac{1}{2} \left( C^{a} D^{b} +
\frac{4(C^2+D^2)}{(2-a-b)^2} \right)
\beta_0^{\frac{a+b+2}{a+b-2}}. \label{monenergy}
\end{equation}
Similarly to the case of the electric solution discussed above, the scalar fields
surrounding the magnetic monopole possess dilaton charges, which
take the values:
\begin{equation}
D_{\phi}= \frac{2}{2-a-b} C \beta_0^{\frac{a+b}{a+b-2}}
\label{adchargemon1}
\end{equation}
and
\begin{equation}
D_{\psi}= \frac{2}{2-a-b} D \beta_0^{\frac{a+b}{a+b-2}} .
\label{adchargemon2}
\end{equation}
For negative value of the parameter $\beta_0$ in the magnetic
solutions (\ref{solmon2}), (\ref{solmon3}) we find the magnetic
counterpart of the electric black hole configuration presented
previously.

It is widely accepted in the literature that finite energy
magnetic solutions correspond to glueballs (see e.g. \cite{kerner})
- effective particles in the low energy sector of gluodynamics.
As we have shown the model (\ref{model}) provides the whole family of
glueball-like states.

In addition there is a solution with infinite energy:
\begin{equation}
\phi (r)=C \Lambda \left( \frac{1}{r\Lambda}
\right)^{\frac{2}{2-a-b}} \label{solmon4}
\end{equation}
\begin{equation}
\psi (r)=D \Lambda \left( \frac{1}{r\Lambda}
\right)^{\frac{2}{2-a-b}} \label{solmon5}
\end{equation}
for $a+b \neq 2$. One can check that, for $a+b >2$ energy density
diverges as  $\sim r^{\delta }$ at the spatial infinity
\begin{equation}
\epsilon = \frac{1}{2} \left( C^{a} D^{b} +
\frac{4(C^2+D^2)}{(2-a-b)^2} \right) \frac{1}{r^4} \left(
\frac{1}{r\Lambda } \right)^{\frac{2a+2b}{2-a-b}}.
\label{monendensing}
\end{equation}
Here $\delta $ takes values from 0 $(a+b \rightarrow \infty )$ to
infinity (when $a+b \rightarrow 2$).

Interestingly enough the electric as well as magnetic solutions of
the model (\ref{model1}) revel an universal property. The
solutions describing scalars are (up to a multiplicative constant)
identical, even though these fields differently couple to the gauge field.
i.e. with $a \neq b$. The energy
density of the fields generated by the electric (magnetic) source
behaves as if there has been only one (effective) scalar field
$\sigma $ coupled to $F^{c \mu \nu}F^{c}_{\mu \nu}$ invariant by $\sigma^{a+b}$.

To summarize, there are three sectors of solutions for the model
defined by (\ref{model1}): screening, confining and glueball-like.
They appear simultaneously i.e. for fixed parameter of the model
one can derive finite as well as infinite energy solution. Because
of the fact that the main aim of such models is to provide
effective description of the low energy features of QCD it is
important to get rid of these non-physical screening solutions. It
can be easily done by adding of the potential term
(\ref{potential}). Unfortunately, it removes not only the finite
energy electric solutions but also all the glueball-like
solutions. This is highly unsatisfactory. The good candidate for
an effective model describing the low energy QCD should model
confinement and possess, at least one, glueball solution.
\section{\bf{The massive scalar fields}}
\label{massec}
In this section we turn to massive scalar fields. In effective
models for the low energy gluodynamics such massive scalars are
usually interpreted as scalar (effective) glueball or/and meson
fields. It is unlikely in the massless case where scalar fields do
not have any clear particle interpretation. Then the Lagrangian
(\ref{model}) takes the following form
\begin{equation}
L=-\frac{1}{4} \left( \frac{\phi}{\Lambda } \right)^a \left(
\frac{\psi}{\Lambda } \right)^b F^{c \mu \nu }F_{\mu \nu }^c
+\frac{1}{2}
\partial_{\mu } \phi \partial^{\mu } \phi +\frac{1}{2}
\partial_{\mu } \psi \partial^{\mu } \psi  -\frac{1}{2} m_{\phi}^2 \phi^2
- \frac{1}{2} m^2_{\psi} \psi^2, \label{model2}
\end{equation}
where $m_{\phi }$, $m_{\psi }$ are masses of the scalars. Let us
now consider the Coulomb problem for the massive model
(\ref{model2}) and compare it with massless solutions. The
pertinent equations of motion can be written as follows
\begin{equation}
\left[ r^2 \left( \frac{\phi}{\Lambda } \right)^a \left(
\frac{\psi}{\Lambda } \right)^b E \right]'=4\pi q \delta(r)
\label{eqmotmass1}
\end{equation}
and
\begin{equation}
\nabla^2_r \phi = -\frac{aE^2}{2\Lambda} \left(
\frac{\phi}{\Lambda } \right)^{a-1} \left(\frac{\psi}{\Lambda }
\right)^b +m_{\phi}^2 \phi \label{eqmotmass2}
\end{equation}
\begin{equation}
\nabla^2_r \psi = -\frac{bE^2}{2\Lambda} \left(
\frac{\phi}{\Lambda } \right)^{a}
 \left(\frac{\psi}{\Lambda } \right)^{b-1} +m_{\psi}^2 \psi.
\label{eqmotmass3}
\end{equation}
Obviously, the additional massive term for scalars does not
inflect the Gauss law. Thus, the electric field is given in
terms of the scalar fields in the following way
\begin{equation}
E = \frac{q}{r^2} \left( \frac{\phi}{\Lambda } \right)^{-a}
 \left(\frac{\psi}{\Lambda } \right)^{-b}.
\label{efieldmass}
\end{equation}
Then remaining equations read
\begin{equation}
\nabla^2_r \phi=-\frac{aq^2}{2 r^4\Lambda} \left(
\frac{\phi}{\Lambda } \right)^{-a-1}
 \left(\frac{\psi}{\Lambda } \right)^{-b} +m_{\phi}^2 \phi
\label{eqmotmass4}
\end{equation}
and
\begin{equation}
\nabla_r^2 \psi=-\frac{bq^2}{2 r^4\Lambda} \left(
\frac{\phi}{\Lambda } \right)^{-a}
 \left(\frac{\psi}{\Lambda } \right)^{-b-1} +m_{\psi}^2 \psi
\label{eqmotmass5}
\end{equation}
Unfortunately, we cannot solve this set of differential equations
analytically. However, because of the fact that model
(\ref{model1}) is usually considered in the connection with low
energy QCD we are mainly interested in the long range behavior of
the solutions. In fact, the asymptotic form of the solution for
$r \rightarrow \infty $ is found to be
\begin{equation}
\phi (r)= G_{\phi} \left(\frac{1}{r} \right)^{\frac{4}{2+a+b}}
\label{solmass1}
\end{equation}
and
\begin{equation}
\psi (r)= G_{\psi} \left(\frac{1}{r} \right)^{\frac{4}{2+a+b}},
\label{solmass2}
\end{equation}
where constants read
\begin{equation}
G_{\phi}=\left( \Lambda^{a+b-1} \frac{aq^2}{2m^2_{\phi}} \left(
\frac{b}{a} \frac{m^2_{\phi}}{m^2_{\psi}} \right)^{-\frac{b}{2}}
\right)^{\frac{1}{2+a+b}} \label{stalaG1}
\end{equation}
and
\begin{equation}
G_{\psi}=\left( \Lambda^{a+b-1} \frac{bq^2}{2m^2_{\psi}} \left(
\frac{a}{b} \frac{m^2_{\psi}}{m^2_{\phi}} \right)^{-\frac{a}{2}}
\right)^{\frac{1}{2+a+b}}.
\label{stalaG2}
\end{equation}
Using (\ref{efieldmass}) we easily obtain the resulting electric field
\begin{equation}
E(r)=q \frac{G^{-a}_{\phi} G_{\psi}^{-b}}{\Lambda^{-a-b}}
\left(\frac{1}{r^2} \right)^{\frac{2-a-b}{2+a+b}} \label{solmass3}
\end{equation}
and the corresponding electric potential
\begin{equation}
U(r)= q \frac{2+a+b}{3(a+b)-2} \frac{G^{-a}_{\phi}
G_{\psi}^{-b}}{\Lambda^{-a-b}} \left(\frac{1}{r}
\right)^{\frac{2-3a-3b}{2+a+b}}. \label{solmass4}
\end{equation}
One can repeat the previous calculations and conclude that this
configuration of fields has infinite energy. For $a+b>\frac{2}{3}$
the total energy diverges due to the behavior of the fields at the
spatial infinity. Thus, the model (\ref{model1}) simulates
the confinement of external electric source, where the
confining potential is given by (\ref{solmass4}). As it was
mentioned before, this potential should not depend on $r$ stronger
than linearly. This provides an upper bound for our parameters:
 $a+b<2$. Finally, the massive scalar model describes
confinement for the following parameters
\begin{equation}
a+b \in \left[\frac{2}{3}, 2 \right]. \label{warunekmass}
\end{equation}
In the very special case, when $a+b=2$ and masses of the scalars
are identical $m_{\phi}^2=m_{\psi}^2=m^2$, we are able to find the
generalized Dick analytical solution
\begin{equation}
\phi=\frac{1}{r} \sqrt{\sqrt{\frac{a}{2}} \frac{q}{m}
+\left(\beta_0^2 - \sqrt{\frac{a}{2}} \frac{q}{m} \right)
e^{-2mr}} \label{analiticmass1}
\end{equation}
and
\begin{equation}
\psi=\frac{1}{r} \left( \frac{b}{a} \right)^{\frac{1}{2+b}}
\sqrt{\sqrt{\frac{a}{2}} \frac{q}{m} +\left(\beta_0^2 -
\sqrt{\frac{a}{2}} \frac{q}{m} \right) e^{-2mr}}.
\label{analiticmass2}
\end{equation}
In agreement with (\ref{warunekmass}) such form of the scalar
fields guarantees the linear dependence of the electric potential
at the large distance. In fact, the electric potential reads
\begin{equation}
U=\frac{1}{q} \sqrt{\frac{2}{a}} \left( \frac{a}{b}
\right)^{\frac{b}{b+2}} \ln \left( e^{2mr}-1+\beta_0^2 \frac{m}{q}
\sqrt{\frac{2}{a}}
 \right), \label{nanaliticpot}
\end{equation}
and for $r \rightarrow \infty$ it diverges linearly.

In the paper \cite{dick2} Dick and Fulchert have proposed a model
where the single scalar field $\phi $ representing the lightest
$0^{++}$ glueball is linearly coupled to the gauge fields i.e.
$\phi F^a_{\mu \nu } F^{a \mu \nu}$. It corresponds to $a=1$ and
$b=0$ in our case. This particular form of the coupling has been
chosen in analogy to the chiral quark model \cite{chiral}. The
confining potential derived from their model reads
$r^{\frac{1}{3}}$. However, one can notice there is another
particle which is relevant in the low energy regime -- a scalar
meson. It seems to be reasonable to assume that both scalars
inflect the low energy dynamics. Due to that one should consider
these scalar fields together. Then the natural generalization of
the Dick--Fulchert model is a theory with one glueball--meson
coupling, that is $\; \phi \psi F^a_{\mu \nu } F^{a \mu \nu }$
(a=1, b=1 in our model). It is very interesting that this form of
interaction ensures also the linear electric potential. This
unexpected and striking result can suggest that the correct low
energy effective action should consist of more than only one
scalar field.
\section{\bf{The non-dynamical scalar fields}}
\label{nondynamical}
In the previous sections we observed that the dynamical scalar fields (massive as
well as massless) can strongly modify electric and magnetic
solutions. Now, we will show that these fields can play very
non-trivial role even when we would drop the kinetic term for
scalar \cite{thooft}. As it will be clarified later a model with
non-dynamical scalar fields can be also interesting in
the context of the low energy gluodynamics.

Let us start with the mixed case: dynamical field $\phi $ and
non-dynamical $\psi $. Additionally, we assume that the potential
for the non-dynamical field is analogous to that in the formula
(\ref{potential}). Then the Lagrangian (in the simplest version)
reads
\begin{equation}
L=-\frac{1}{4} \left( \frac{\phi}{\Lambda } \right)^a \left(
\frac{\psi}{\Lambda } \right)^b F^{c \mu \nu }F_{\mu \nu }^c
+\frac{1}{2}
\partial_{\mu } \phi \partial^{\mu } \phi - \alpha \Lambda^4 \left(
\frac{\psi }{\Lambda } \right)^{d}, \label{model3}
\end{equation}
where $d$ is a positive parameter. We will analyze the Coulomb
problem. Then the pertinent equations of motion take the following
form
\begin{equation}
\left[ r^2 \left( \frac{\phi}{\Lambda } \right)^a \left(
\frac{\psi}{\Lambda } \right)^b E \right]'=4\pi q \delta(r)
\label{eqmotndyn1}
\end{equation}
and
\begin{equation}
\nabla^2_r \phi = -\frac{aE^2}{2\Lambda} \left(
\frac{\phi}{\Lambda } \right)^{a-1} \left(\frac{\psi}{\Lambda }
\right)^{b}, \label{eqmotndyn2}
\end{equation}
\begin{equation}
\alpha d \Lambda^3 \left( \frac{\psi }{\Lambda } \right)^{d-1} =
\frac{bE^2}{2\Lambda} \left( \frac{\phi}{\Lambda } \right)^{a}
 \left(\frac{\psi}{\Lambda } \right)^{b-1}.
\label{eqmotndyn3}
\end{equation}
Obviously, the variation with respect to the non-dynamical field
gives a constrain on the scalar field $\psi$ (\ref{eqmotndyn3}).
We can solve it and express $\psi $ in terms of the dynamical
fields $\phi $ and $E$
\begin{equation}
\left( \frac{\psi }{\Lambda } \right) = \left( \frac{b E^2 }{2
\alpha d \Lambda^4} \left( \frac{\phi}{\Lambda} \right)^a
\right)^{\frac{1}{d-b}}. \label{psisol}
\end{equation}
Then the remaining field equations
\begin{equation}
\left[ r^2 \left( \frac{\phi}{\Lambda } \right)^{\frac{a d}{d-b}}
\left( \frac{bE^2}{2 \alpha d \Lambda^4} \right)^{\frac{b}{d-b}} E
\right]'=4\pi q \delta(r) \label{eqmotndyn4}
\end{equation}
and
\begin{equation}
\nabla^2_r \phi = -\frac{aE^2}{2\Lambda} \left(
\frac{bE^2}{2\alpha d \Lambda^4} \right)^{\frac{b}{d-b}} \left(
\frac{\phi}{\Lambda } \right)^{\frac{ad}{d-b}-1}.
 \label{eqmotndyn5}
\end{equation}
Finally, we obtain the following solutions
\begin{equation}
\phi (r)=H \Lambda \left( \frac{1}{r\Lambda }
\right)^{\frac{4d}{a d + 2(d+b)}} \label{solndyn1}
\end{equation}
and
\begin{equation}
E(r) = \Lambda^2 q^{\frac{d-b}{d+b}} \left( \frac{b}{2\alpha d}
\right)^{-\frac{b}{d+b}} H^{-\frac{ad}{d+b}} \left(
\frac{1}{r\Lambda } \right)^{\frac{4(d-b)}{2(d+b)+ad}}.
\label{solndyn2}
\end{equation}
Here constant $H$ takes the form
\begin{equation}
H=q^{\frac{2d}{ad+2(d+b)}} \left(\frac{b}{2\alpha d}
\right)^{-\frac{b}{ad+2(d+b)}} \left( \frac{a}{4}
\frac{(ad+2(d+b))^2}{(d-b)(ad+4b)}
\right)^{\frac{b+d}{ad+2(d+b)}}.
\end{equation}
The corresponding electric potential reads
\begin{equation}
U(r)=\Lambda q^{\frac{d-b}{d+b}} \frac{ad+2(d+b)}{ad-2(d+b)}
\left( \frac{b}{2\alpha d} \right)^{-\frac{b}{d+b}}
H^{-\frac{ad}{d+b}} \left( \frac{1}{r\Lambda }
\right)^{\frac{ad-2(d+b)}{ad+2(d+b)}}. \label{solndyn3}
\end{equation}
It has confining-like behavior if the parameters fulfill the following
condition
\begin{equation}
1>\frac{2(d+b)-ad}{2(d+b)+ad} >0. \label{nonequal}
\end{equation}
On the other hand, because of the fact that the non-dynamical
field play role of the Lagrange multiplier, the action can be
rewritten in terms of the dynamical fields only. The new
Lagrangian has the form
\begin{equation}
L=-\frac{1}{4} \left( \frac{\phi}{\Lambda }
\right)^{\frac{ad}{d-b}} \left( \frac{F}{\Lambda }
\right)^{\frac{b}{d-b}} F^{c \mu \nu }F_{\mu \nu }^c +\frac{1}{2}
\partial_{\mu } \phi \partial^{\mu } \phi,  \label{modeleqv}
\end{equation}
where $F=\frac{1}{2} F^{c \mu \nu}F^c_{\mu \nu }$. One can easily
check that, in the case of Coulomb problem, the pertinent
equations of motion are identical to (\ref{eqmotndyn4}) and
(\ref{eqmotndyn5}). The system with the dielectric function
governed by dynamical and non-dynamical scalar field is equivalent
to the model where the permittivity depends on the dynamical
scalar as well as gauge fields. Thus, we have obtained a
generalized version of an effective theory proposed by Pagels and
Tomboulis a long time ago \cite{pagels}. On the contrary to the
original version (\ref{modeleqv}) contains non-minimal
gauge--scalar coupling along with an exponent of the standard
Yang--Mills invariant. This equivalency seems to be particularly
interesting since the Pagels--Tomboulis model (and the similar
Adler--Savvidy model \cite{savvidy}, \cite{adler},
\cite{mendel1}), reproduces the trace anomaly already on the
classical level and gives correct prediction for the confining
potential.

Obviously, it is possible to obtain the standard Pagels--Tomboulis
model treating also the second scalar field as non-dynamical. Let us
assume that the potential for this field is
$$V(\phi) = \alpha' \left(\frac{\phi}{\Lambda} \right)^{e},$$
where $e$ is a new parameter and $\alpha'$ is a dimensionless
constant. Then both scalars can be written as functions of the
electric field. Finally, we obtain
\begin{equation}
L_{PT}=-\frac{1}{4} \left( \frac{F}{\Lambda^4}
\right)^{2\delta}F^c_{\mu \nu}F^{c \mu \nu } \label{pagels}
\end{equation}
where
$$ 2\delta =\frac{ad+be}{ed-ad-be}. $$
As it was shown in \cite{pagels}, \cite{my2}, this model
guarantees the confinement of external, Abelian electric sources
for
\begin{equation}
\delta \geq \frac{1}{4}. \label{warunekPT}
\end{equation}
Duality demonstrated above allows us to believe that an ultimate gluodynamics
effective theory would share properties correctly described by all instances
presented in this paper.
\section{\bf{Summary}}
The main aim of this work has been analyzing of electric and
magnetic solutions in the standard classical Yang--Mills theory
coupled to two scalar fields. For particular values of the
parameters of the discussed models (with massless, massive as well
as non-dynamical scalar fields) we have obtained confining-like
field configurations. External, static, electric sources have
infinite energy due to the long range behavior of the fields. From
this point of view all models are very similar. Clearly, the
values of the parameters are model dependent and different
potentials known from fits to the phenomenological data are
derived for different $a$ and $b$. One can also notice that the
model with massless scalars is not able to describe the linear
electric potential which corresponds to the limit $a,b \rightarrow
\infty $. In the other words this model can only approximate the
linear potential (but with arbitrary accuracy). Moreover, in
contradiction to other Lagrangians, there are finite energy
electric solutions in the massless scalar model. Here, such
solutions representing screening phenomena at the classical level,
are rather non-physical. They are removed from the spectrum of the
theory by adding a potential term for the massless fields.

In spite of the fact that presented models in the similar way
describe the confining solutions they differ profoundly.

The main difference between our models has its source in the
physical interpretation of the scalars. This is strongly connected
with the glueball problem. Let us firstly discuss the simplest
case i.e. model with the massive scalar fields. Then these field
can be identified with some massive particles relevant to the low
energy sector of QCD. Evident candidates are the scalar glueball
$0^{++}$ and a scalar meson. Both objects exist due to the
non-perturbative effect, confinement of the gauge and quark fields
respectively. Following that one could expect that these field
should appear in the 'democratic' manner. They should couple with
the gauge field identically: $a=b$. This gives us prescription how
correctly generalize QCD motivated Dick--Fulcher model. Namely,
instead of one glueball coupling we introduce one glueball--meson
coupling. Surprisingly, such form of the dielectric function
provides the standard linear electric potential. For identical
masses of the scalar glueball and meson the analytical solution
has been explicitly found.

In case of the massless fields situation is a little bit more
subtle. Massless scalars cannot be rather identified with any
particles. There are not massless particles in the low energy
sector of QCD. Even thought scalars do not possess particle-like
interpretation the existence of the glueballs is not excluded.
There are at least two possible ways of introducing glueballs in
the framework of massless scalar model. Firstly, they can appear
as finite energy, magnetic solutions. However, such configurations
exist simultaneously with electric screening solutions. If we get
rid of the screening solutions by adding a potential term to the
Lagrangian then also finite energy, magnetic solutions disappear.
Unfortunately, we are not able to preserve this glueball-like
sector and remove the screening sector. On the other hand one
could try to find glueballs spectrum in the analogous way as it
has been done for charmonium and bottomium states i.e. by means of
Schr\"{o}dinger equation. In this picture glueball would consist
of magnetic monopole and anti-monopole. Then the potential in
Schr\"{o}dinger equation would be potential between magnetic
monopoles \cite{my3}. Similar glueball model has been recently
considered in the paper \cite{glueball}.

Even more interesting from glueball point of view is the model
with the non-dynamical scalar field (or fields). Such model
corresponds to the generalized Pagels--Tomboulis theory. Here
glueballs could appear as toroidal solitons with nontrivial Hopf
index \cite{niemi}. It follows from the observation that the
restricted version of Pagels--Tomboulis model with $\delta
=-\frac{1}{4}$ possesses such solutions \cite{aratyn} (but then
electric sources are not confined). We expect that additional
scalar field could joint the glueball and confining sectors.

To conclude, all models presented here can be treated as
candidates for the effective model for the low energy QCD. They
describe confinement and give reasonable quark--anti-quark
potentials. However, the problem of the
glueballs has  not been  satisfactorily solved yet. We would 
like to analyze it in our forthcoming paper.

\end{document}